\documentclass[conference]{IEEEtran}
\IEEEoverridecommandlockouts

\usepackage{cite}
\usepackage{amsmath,amssymb,amsfonts}
\usepackage{algorithm}
\usepackage{algorithmic}
\usepackage{graphicx}
\usepackage{textcomp}
\usepackage{xcolor}
\def\BibTeX{{\rm B\kern-.05em{\sc i\kern-.025em b}\kern-.08em
    T\kern-.1667em\lower.7ex\hbox{E}\kern-.125emX}}

\begin{document}
\title{LLM-Assisted LDPC Decoding via Syndrome-Verified Semantic Priors\thanks{\textit{(Corresponding author: Hyun Jong Yang)}}}

\author{\IEEEauthorblockN{Sojeong Park}
\IEEEauthorblockA{\textit{Dept. of Electrical Engineering} \\
\textit{Pohang University of Science and Technology}\\
Pohang, Republic of Korea \\
sojeong@postech.ac.kr}
\and
\IEEEauthorblockN{Hyeonsu Lyu\IEEEauthorrefmark{1}, Jaehyun Choi\IEEEauthorrefmark{2}, and Hyun Jong Yang\IEEEauthorrefmark{1}\IEEEauthorrefmark{2}}
\IEEEauthorblockA{\textit{\IEEEauthorrefmark{1}Institute of New Media and Communications} \\
\textit{\IEEEauthorrefmark{2}Dept. of Electrical and Computer Engineering} \\
\textit{Seoul National University}\\
Seoul, Republic of Korea \\
\{hs.lyu, jhchoi0226, hjyang\}@snu.ac.kr}
}
\maketitle
\bstctlcite{IEEEexample:BSTcontrol}
\begin{abstract}
Semantic communication exploits the meaning of the payload, which bit-level processing discards. When channel decoding fails on a natural language payload, the errors appear as corrupted characters in the recovered text. A large language model (LLM) infers the intended characters from the semantic context, but it can also produce incorrect corrections. Applying them directly introduces new bit errors when the LLM modifies characters incorrectly. In this paper, we propose an LLM-assisted decoding framework for low-density parity-check (LDPC) codes. Rather than trusting LLM predictions, the decoder evaluates the modified characters jointly against the parity-check constraints and admits only the accepted corrections as verified semantic priors. These priors are injected as soft updates to the channel log-likelihood ratios, preserving the original channel evidence without modifying the decoder. A subsequent belief propagation pass distributes the injected evidence across the check nodes, recovering not only the injected bits but also the residual errors that the LLM fails to correct. Simulations demonstrate a 73\% bit error rate reduction over a conventional decoder at 2.0 dB, whereas doubling its iterations to the same budget yields only 21\%. The verification maintains an injection precision above 0.88 despite inaccurate LLM predictions, demonstrating that semantic knowledge can be reliably translated into physical-layer reliability gains.
\end{abstract}

\begin{IEEEkeywords}
Semantic communication, large language model, channel decoding, LDPC codes, belief propagation.
\end{IEEEkeywords}

\section{Introduction}
Natural language now accounts for a growing share of wireless traffic, driven by the wide deployment of large language model (LLM)-based services. These text payloads possess rich contextual structures. Conventional receivers, processing data purely as independent sequences of bits, leave this semantic information entirely unexploited. Leveraging this inherent structure forms the central premise of semantic communications, establishing it as a pivotal research direction for next-generation networks \cite{luo2022semantic, shao2024theory, yang2022semantic}.

Early studies in this area replace the conventional transceiver with an end-to-end learned encoder, shifting the design objective from bit-level accuracy to semantic fidelity \cite{xie2021deep, peng2024robust, huang2022toward}. Utilizing the extensive prior knowledge of LLMs, recent systems employ these models for semantic extraction and recovery. Such approaches either fuse LLM-generated descriptions with visual features \cite{zhao2024lamosc} or utilize an LLM as a shared knowledge base for multimodal transmissions \cite{jiang2024large}. Building on the concept of semantic recovery, characters deliberately omitted at the transmitter can be reconstructed at the receiver \cite{park2026robust}. These designs operate at the source level, isolating the payload semantics from the physical layer. A complementary approach preserves the standard protocol stack and injects semantic information directly into physical-layer processing. Our prior work follows this paradigm, leveraging LLM-corrected text to identify reliable symbols for data-aided channel estimation \cite{park2026semantic, park2026semanticmimo}. These designs operate on uncoded symbols, leaving the channel code structure unexploited.

Low-density parity-check (LDPC) codes~\cite{gallager1962low}, decoded via belief propagation (BP), are widely deployed in modern wireless standards~\cite{richardson2018design,3gpp38212}. However, BP frequently fails to converge to a valid codeword. Under finite block lengths and limited iteration budgets, the decoder often enters a trapping set~\cite{richardson2003error}. Within these sets, parity checks connected to an even number of erroneous bits remain satisfied, concealing the underlying errors. We refer to such checks as deceptively satisfied. The few remaining unsatisfied checks fail to localize the error pattern. The syndrome weight stops decreasing, additional iterations extract no new parity information, and the decoder settles into a fixed state responsible for the error floor~\cite{dolecek2010analysis,butler2014error}. Escaping this convergence failure requires evidence external to the code. However, existing remedies still draw that evidence from within the decoder, biasing or flipping the bits of the dominant trapping sets identified during or prior to decoding~\cite{han2009low,kang2011iterative}. For a natural language payload, a code-independent source of evidence is instead available. Residual bit errors corrupt individual characters of the recovered text, and the intended characters can be inferred from the surrounding context.

\begin{figure*}[t]
    \centering
    \includegraphics[width=16cm]{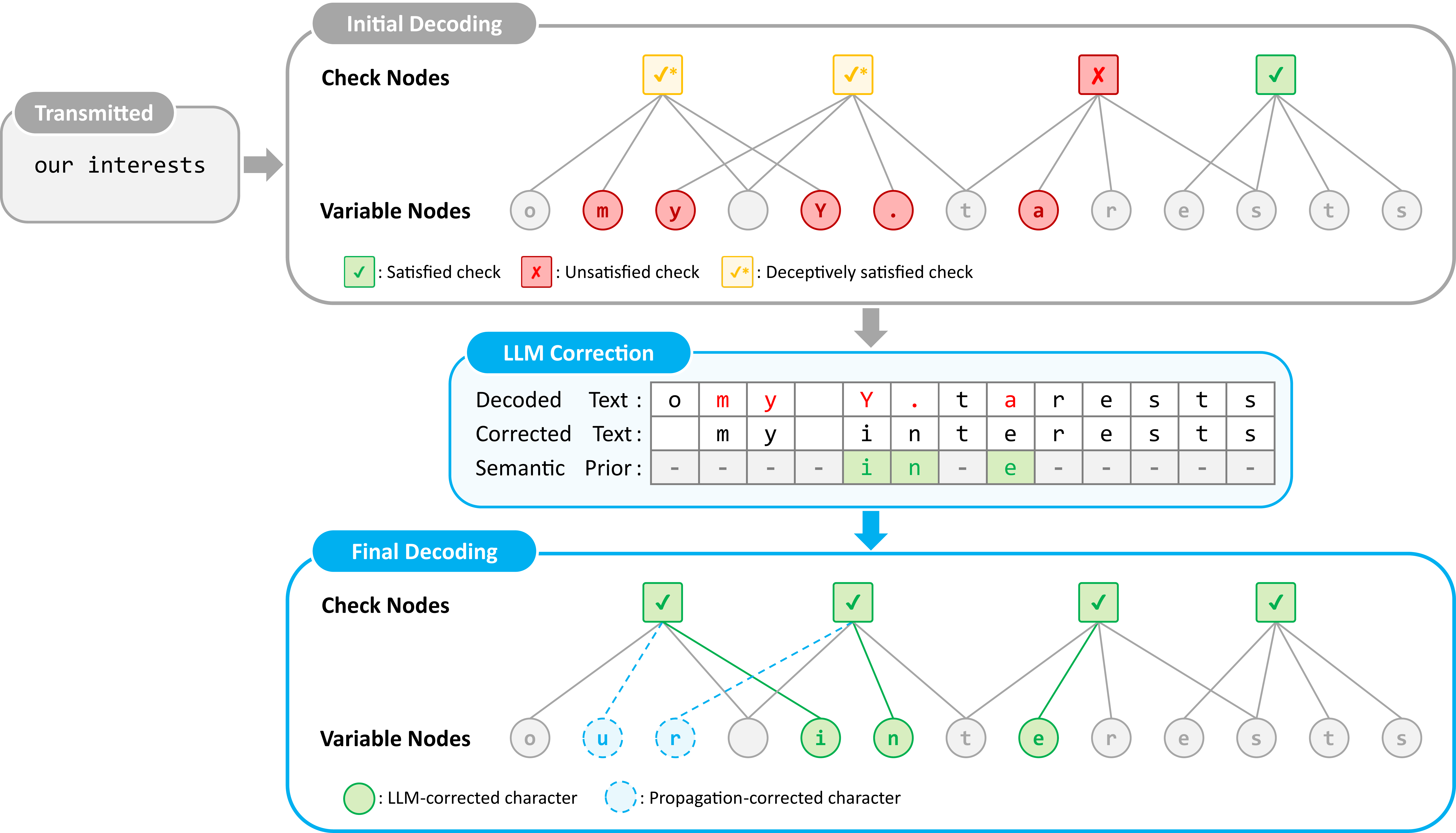}
    \caption{Proposed LLM-assisted LDPC decoding with verified semantic prior injection.}
    \vspace{-2mm}
    \label{fig:system_model}
\end{figure*}

Recent studies have integrated LLMs into channel decoding. For convolutional codes, \cite{li2026llm} fuses linguistic scores with channel metrics to prune candidate paths during Viterbi decoding. For short block codes, \cite{hao2026semantic} generates multiple LLM reconstructions post-decoding, selecting the candidate that minimizes the weighted Hamming distance to the received signal. In both cases, a correction only affects the positions the LLM modifies and cannot spread to the remaining error bits. LDPC codes decoded by BP remove this limitation, as message passing propagates a correction from one bit to every other bit sharing a check. A concurrent effort \cite{thai2026soft} operates on LDPC codes, admitting LLM suggestions by their multi-sample voting confidence, which reflects self-consistency and leaves confident errors undetected. We instead admit a correction only when the corrected word stays consistent with the parity checks, making acceptance a code-level test rather than a model-confidence criterion. This criterion protects the decoder from confident LLM hallucinations.

In this paper, we propose a semantic-aware LDPC decoding framework that admits semantic evidence only after verifying it against the syndrome of the code. Fig.~\ref{fig:system_model} illustrates the procedure. The transmitted text is \texttt{our interests}. The initial BP pass terminates with a nonzero syndrome and returns \texttt{omy Y.tarests}, where some errors hide behind deceptively satisfied checks. The LLM restores the corrupted word to \texttt{interests}. Simultaneously, it inaccurately rewrites \texttt{our} as \texttt{my}, shortening the sequence. Following the approach in \cite{park2026semanticmimo}, the Needleman-Wunsch (NW) algorithm \cite{needleman1970general} aligns the corrected text to the decoded output. This alignment prevents length mismatches from displacing the remaining corrections from their coded bit positions. The decoder cross-examines the candidate corrections jointly against the parity-check structure. In this example, the characters \texttt{i}, \texttt{n}, and \texttt{e} pass the test to form the semantic prior, while the incorrect edit on \texttt{our} is rejected outright. The system converts the accepted characters into a priori log-likelihood ratios (LLRs), injecting them strictly at bits with small posterior magnitudes. A subsequent BP pass diffuses this soft evidence through the check nodes. This propagation successfully recovers the full phrase \texttt{our interests} by restoring \texttt{u} and \texttt{r}, characters overlooked by the LLM. Simulations over an additive white Gaussian noise (AWGN) channel confirm the effectiveness of the proposed method. The verified injection consistently outperforms both the conventional BP decoder under an equal iteration budget and unverified variants that blindly trust LLM outputs. Since it modifies only the input LLRs, the framework preserves the BP algorithm and can be applied to any deployed LDPC decoder.

\section{System Model}
This paper considers a text transmission system where a natural-language payload is protected by channel coding and recovered via iterative decoding at the receiver. When decoding fails, the recovered text carries corrupted characters that an LLM can identify from the surrounding context. The LLM corrects these characters and returns a modified text. We filter unreliable LLM outputs through a verification step, injecting only accepted corrections as a priori LLRs into the decoder. The injected priors bias the corresponding bits toward the corrected characters, and a subsequent BP pass propagates this evidence to resolve remaining errors. Fig.~\ref{fig:system_model} shows an overview of the proposed LLM-assisted decoding framework.

\subsection{Transmission Model}
The transmitter conveys a text payload of $M$ characters,\begin{equation}
\mathbf{t} = \bigl[\,t^{(1)},\, t^{(2)},\, \dots,\, t^{(M)}\,\bigr],
\label{eq:text}
\end{equation}
with each character mapped to $q$ bits. The payload forms an information bit vector
$\mathbf{u}\in\{0,1\}^{k}$. The bits of $t^{(i)}$ form the index set
\begin{equation}
\mathcal{B}_i = \bigl\{\,q(i-1)+1,\; q(i-1)+2,\; \dots,\; qi\,\bigr\},
\; i = 1,\dots,M .
\label{eq:bitmap}
\end{equation}

The information vector is encoded by a systematic $(n,k)$ LDPC code with parity-check matrix $\mathbf{H}\in\{0,1\}^{m\times n}$, where $m$ denotes the number of parity checks, yielding the codeword
\begin{equation}
\mathbf{c} = [\,c_1,\dots,c_n\,] = [\,\mathbf{u},\, \mathbf{p}\,]\in\{0,1\}^{n},
\label{eq:codeword}
\end{equation}
where $c_j$ is the $j$-th coded bit and $\mathbf{p}$ denotes the parity bits. The codeword is transmitted over a noisy channel, and the receiver observes $\mathbf{y} = [y_1,\dots,y_n]$, where $y_j$ is the channel output corresponding to the coded bit $c_j$. From $\mathbf{y}$, the decoder computes the per-bit channel LLRs
\begin{equation}
L_{\mathrm{ch}}(j)
= \ln\frac{p\bigl(y_j \mid c_j = 0\bigr)}{p\bigl(y_j \mid c_j = 1\bigr)},
\qquad j = 1,\dots,n .
\label{eq:llr}
\end{equation}
The specific channel realization is given in Section~\ref{sec:sim}. These values form the channel LLR vector $\mathbf{L}_{\mathrm{ch}} = \bigl[\,L_{\mathrm{ch}}(1),\, \dots,\, L_{\mathrm{ch}}(n)\,\bigr]$.

 \begin{figure} [t]
    \centering
    \includegraphics[width=0.97\linewidth]{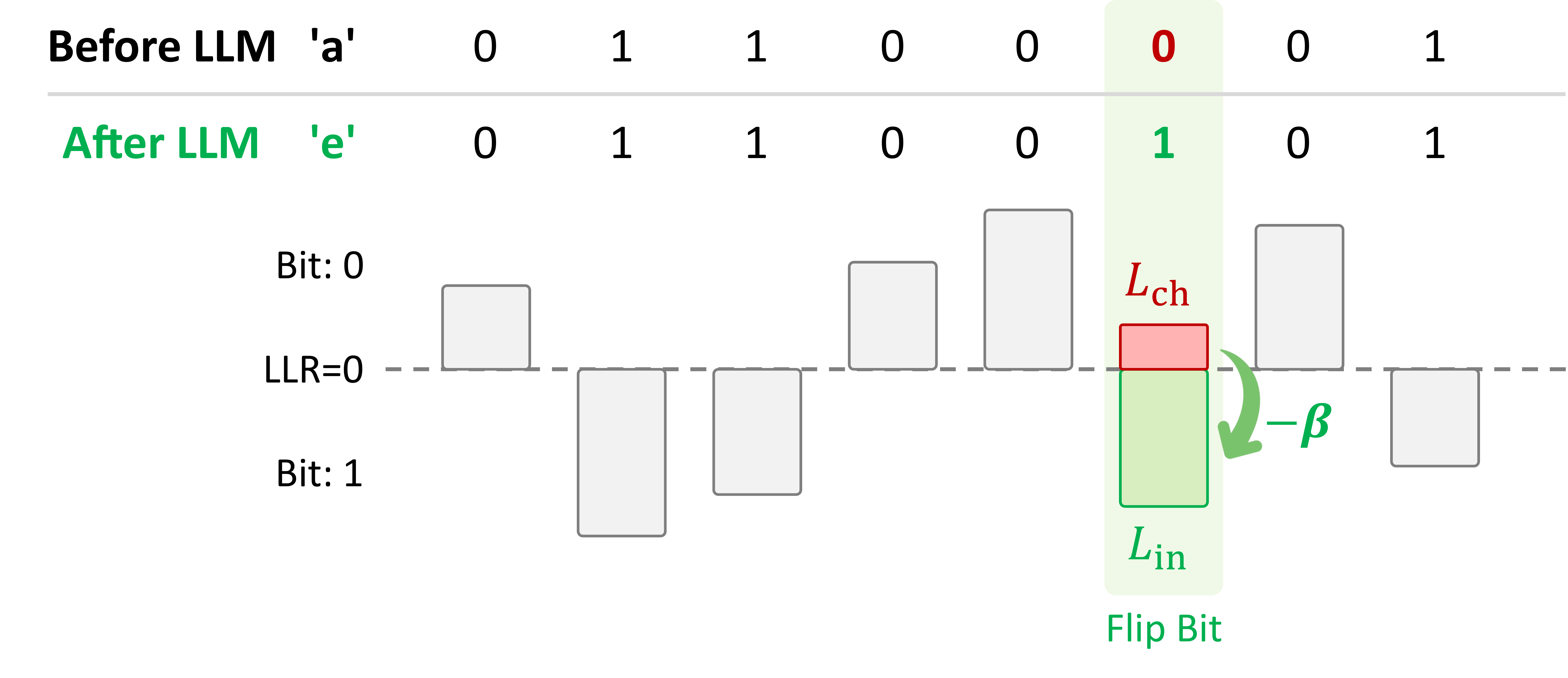}
    \caption{Illustration of the LLR injection rule for an accepted character correction.}
    \vspace{-2mm}
    \label{fig:llr_injection}
\end{figure}

\begin{algorithm}[t]
\vspace*{2mm}
\caption{Proposed LLM-Assisted Decoding}
\label{alg:main}
\begin{algorithmic}[1]
\REQUIRE $\mathbf{L}_{\mathrm{ch}}$, $\mathbf{H}$, $I_{\max}$, $\epsilon$, $\theta$, $\beta$
\STATE $\hat{\mathbf{c}}, \mathbf{L}_{\mathrm{post}} \leftarrow \mathrm{BP}(\mathbf{L}_{\mathrm{ch}}, I_{\max})$
\IF{$w(\hat{\mathbf{c}}) \neq 0$}
  \STATE $\hat{\mathbf{t}} \leftarrow$ source decoding of $\hat{\mathbf{c}}$
  \STATE $\hat{\mathbf{t}}_{\mathrm{LLM}} \leftarrow$ LLM correction of $\hat{\mathbf{t}}$
  \STATE $\bar{\mathbf{t}}, \bar{\mathbf{t}}_{\mathrm{LLM}} \leftarrow$ NW alignment of $\hat{\mathbf{t}}$ and $\hat{\mathbf{t}}_{\mathrm{LLM}}$
  \STATE $\mathcal{S} \leftarrow \mathcal{P}$ from \eqref{eq:candidates}
  \WHILE{$w(\mathbf{c}(\mathcal{S})) > w(\hat{\mathbf{c}}) + \epsilon$}
    \STATE compute $\delta_i$ for all $i \in \mathcal{S}$ using \eqref{eq:loo}
    \STATE \textbf{if} $\min_i \delta_i \ge 0$ \textbf{then break}
    \STATE $\mathcal{S} \leftarrow \mathcal{S} \setminus \{\arg\min_i \delta_i\}$
  \ENDWHILE
  \STATE $\mathcal{S}^\star \leftarrow \mathcal{S}$
  \STATE $\mathcal{A} \leftarrow$ bit positions flipped by $\mathcal{S}^\star$
  \STATE $\mathcal{I} \leftarrow \{ j \in \mathcal{A} : |L_{\mathrm{post}}(j)| \le \theta \}$
  \IF{$\mathcal{I} \neq \varnothing$}
    \STATE construct $\mathbf{L}_a$ from $\mathcal{I}$ and $\beta$ using \eqref{eq:La}
    \STATE $\mathbf{L}_{\mathrm{in}} \leftarrow \mathbf{L}_{\mathrm{ch}} + \mathbf{L}_a$
    \STATE $\hat{\mathbf{c}} \leftarrow \mathrm{BP}(\mathbf{L}_{\mathrm{in}}, I_{\max})$
  \ENDIF
\ENDIF
\RETURN $\hat{\mathbf{c}}$
\end{algorithmic}
\end{algorithm}

\subsection{Initial BP Decoding}
The receiver first performs standard sum-product BP over the Tanner graph of $\mathbf{H}$. Each iteration produces a posterior LLR $L_{\mathrm{post}}(j)$ for every coded bit, from which the hard decision $\hat{c}_j = \bigl[\,L_{\mathrm{post}}(j) < 0\,\bigr]$ is formed. The posterior LLRs form the vector $\mathbf{L}_{\mathrm{post}} = \bigl[\,L_{\mathrm{post}}(1),\, \dots,\, L_{\mathrm{post}}(n)\,\bigr]$.
Collecting these bits gives the estimated codeword
$\hat{\mathbf{c}} = [\,\hat{c}_1,\dots,\hat{c}_n\,]$ and the syndrome
\begin{equation}
\mathbf{s} \equiv \mathbf{H}\,\hat{\mathbf{c}}^{\top} \pmod{2}.
\label{eq:syndrome}
\end{equation}
Decoding runs for $I_{\max}$ iterations, and the hard decision at the final iteration is taken as the decoder output. The information part of $\hat{\mathbf{c}}$ is source-decoded into an initial text estimate
\begin{equation}
\hat{\mathbf{t}} = \bigl[\,\hat{t}^{(1)},\, \hat{t}^{(2)},\, \dots,\,
\hat{t}^{(M)}\,\bigr].
\label{eq:decoded_text}
\end{equation}
In this recovered text, any residual bit errors manifest directly as typographical mistakes.

\subsection{LLM-Assisted BP Decoding}
After decoding, residual errors in $\hat{\mathbf{t}}$ fall into two categories. The first category includes errors connected to at least one unsatisfied check. The second category involves errors connected to satisfied checks. With an even number of erroneous bits, these checks remain satisfied. As illustrated in Fig.~\ref{fig:system_model}, such errors hide behind \emph{deceptively satisfied} checks. In both cases, BP has exhausted all available parity information. Consequently, additional iterations yield no further improvement, leaving the decoder trapped in a fixed state.

The proposed method resolves these residual errors by using the payload text as an external source of evidence. An LLM infers the intended characters at the corrupted positions, and the decoder injects the accepted corrections as a priori LLRs into a second BP pass. Section~\ref{sec:proposed} details this procedure.

 \begin{figure*}[t]
    \centering
    \includegraphics[width=\textwidth]{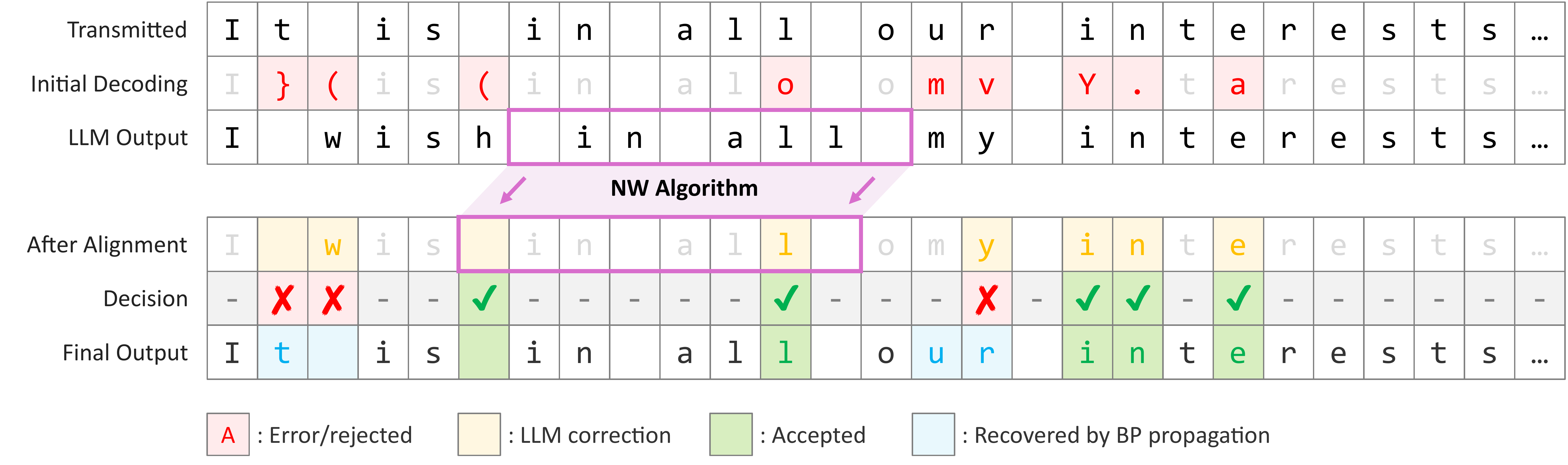}
    \caption{Semantic prior construction for a received frame, showing the sequence alignment, the verification decisions, and the bits recovered by BP propagation.}
    \vspace{-2mm}
    \label{fig:example}
\end{figure*}

\section{Proposed LLM-Assisted Decoding} \label{sec:proposed}

This section presents the construction of $\mathbf{L}_a$. The proposed decoder is invoked when BP terminates with a nonzero syndrome. It consists of three steps: LLM-based text processing, syndrome-based verification, and selective LLR injection. The overall procedure is summarized in Algorithm~\ref{alg:main}.

\subsection{LLM-Based Text Processing}
\label{sec:correction}
The initial estimate $\hat{\mathbf{t}}$ contains corrupted characters at the positions of the residual bit errors. The LLM is prompted to substitute only these characters, with paraphrasing, reordering, insertion, and deletion explicitly disallowed, and returns
\begin{equation}
\hat{\mathbf{t}}_{\mathrm{LLM}} = \bigl[\,\hat{t}^{(1)}_{\mathrm{LLM}},\,
\dots,\, \hat{t}^{(M')}_{\mathrm{LLM}}\,\bigr].
\label{eq:llm_text}
\end{equation}
Although the prompt requires strict character count preservation, the LLM frequently violates this constraint. Generating text token by token with variable character lengths prevents the model from guaranteeing an exact output length. A single-character shift maps the subsequent corrections onto unrelated coded bits, and the resulting prior is injected at unintended positions. Following \cite{park2026semanticmimo}, the NW algorithm \cite{needleman1970general} aligns $\hat{\mathbf{t}}$ and $\hat{\mathbf{t}}_{\mathrm{LLM}}$ into two sequences $\bar{\mathbf{t}}$ and $\bar{\mathbf{t}}_{\mathrm{LLM}}$ of equal length. This alignment preserves the coded bit positions of each corrected character. Comparing the aligned sequences position by position gives the set of correction candidates
\begin{equation}
\mathcal{P} = \bigl\{\, i : \bar{t}^{(i)} \neq
\bar{t}^{(i)}_{\mathrm{LLM}} \,\bigr\}.
\label{eq:candidates}
\end{equation}

\subsection{Syndrome-Based Verification}
\label{sec:verify}
Incorrect candidates in $\mathcal{P}$ introduce false priors that degrade decoding. The framework prevents this by verifying the candidates against the parity-check structure before use. The verification criterion is the syndrome weight
\begin{equation}
w(\mathbf{c}) = \bigl\lVert \mathbf{H}\,\mathbf{c}^{\top} \bmod 2
\bigr\rVert_1
\label{eq:weight}
\end{equation}
which counts the unsatisfied checks. A valid codeword attains $w=0$, and a smaller $w$ indicates a bit vector closer to a valid codeword. 
Since a single parity check involves multiple coded bits, several candidate corrections often affect the same check. The impact of one correction on the syndrome weight therefore depends on the other applied candidates. This interdependence requires the decoder to evaluate all proposed corrections jointly. The verification begins with all candidates applied at once. Let $\mathcal{S}\subseteq\mathcal{P}$ denote the set of candidates that are currently retained, and let $\mathbf{c}(\mathcal{S})$ denote the bit vector obtained by applying them to $\hat{\mathbf{c}}$. The set is initialized as $\mathcal{S}=\mathcal{P}$. If the joint effect of the candidates keeps the syndrome weight within the tolerance,
\begin{equation}
w\bigl(\mathbf{c}(\mathcal{S})\bigr) \le w(\hat{\mathbf{c}}) + \epsilon ,
\label{eq:accept}
\end{equation}
the entire set is accepted. A correct candidate does not always reduce the syndrome weight. Consider two erroneous bits sharing a deceptively satisfied check. Correcting one of them leaves the check unsatisfied, and $w$ increases even though the bit vector has moved closer to the transmitted codeword. A strict decrease criterion would discard exactly the corrections that resolve such hidden errors. The tolerance $\epsilon$ admits these transient increases. Keeping $\epsilon$ small relative to the number of parity checks still rejects a set of largely incorrect candidates.

Otherwise, the candidates that increase the syndrome weight are removed. The contribution of each candidate is quantified by excluding it from the current set,
\begin{equation}
\delta_i = w\bigl(\mathbf{c}(\mathcal{S}\setminus\{i\})\bigr)
- w\bigl(\mathbf{c}(\mathcal{S})\bigr),
\label{eq:loo}
\end{equation}
where a negative $\delta_i$ indicates that the syndrome weight decreases upon the exclusion of candidate $i$. The candidate attaining the smallest $\delta_i$ is discarded, and the procedure repeats until \eqref{eq:accept} is satisfied. The loop terminates early when no single exclusion decreases the weight, which occurs when candidates reduce it only in combination. If every candidate is discarded in sequence, $\mathcal{S}^\star$ is empty and the frame receives no prior. The surviving set $\mathcal{S}^\star$ constitutes the accepted corrections, and $\mathcal{A}$ collects the coded bit positions that they flip.

\subsection{Selective LLR Injection}
\label{sec:inject}
 
The accepted corrections are converted into a priori LLRs and combined with the initial channel LLRs. A subsequent BP pass then operates on these updated values. This approach preserves the original channel evidence, allowing the code constraints to dictate the final bit values.
 
This injection process operates selectively. The magnitude of $L_{\mathrm{post}}(j)$ serves as a reliability metric for bit $j$. To protect highly confident estimates, bits with a large magnitude bypass the injection. Consequently, the LLM prior does not disrupt decisions already established by the channel observations and code constraints. The injection set is accordingly restricted to
\begin{equation}
\mathcal{I} = \bigl\{\, j \in \mathcal{A} :
\bigl\lvert L_{\mathrm{post}}(j) \bigr\rvert \le \theta \,\bigr\},
\label{eq:injset}
\end{equation}
where $\theta$ is a confidence threshold. For $j\in\mathcal{I}$, let $b_j\in\{0,1\}$ denote the bit value indicated by the accepted correction. The a priori LLR is
\begin{equation}
L_a(j) =
\begin{cases}
\beta\,\bigl(1 - 2 b_j\bigr), & j \in \mathcal{I},\\[2pt]
0, & \text{otherwise},
\end{cases}
\label{eq:La}
\end{equation}
where $\beta>0$ controls how strongly the language prior is trusted. The components $L_a(j)$ form the a priori LLR vector $\mathbf{L}_a$, which constitutes the semantic prior injected into the second BP pass. The LLR supplied to the second BP pass is
\begin{equation}
L_{\mathrm{in}}(j) = L_{\mathrm{ch}}(j) + L_a(j),
\qquad j = 1,\dots,n,
\label{eq:Lin}
\end{equation}
and the resulting vector is denoted by $\mathbf{L}_{\mathrm{in}}$. Fig.~\ref{fig:llr_injection} illustrates this rule for an accepted correction. Only the bits changed by the correction receive the new prior, whereas the remaining bits retain their channel LLRs. 
By propagating this new evidence through the check nodes, BP also recovers additional bits that received no direct prior. Fig.~\ref{fig:example} illustrates the complete procedure for a received frame. The LLM produces plausible but incorrect words, and the verification step discards them. Injecting the surviving characters as priors allows the second BP pass to resolve remaining errors and recover the transmitted frame. The decoder therefore exploits semantic evidence without requiring a flawless LLM output.

 \begin{figure} [t]
    \centering
    \includegraphics[width=0.8\linewidth]{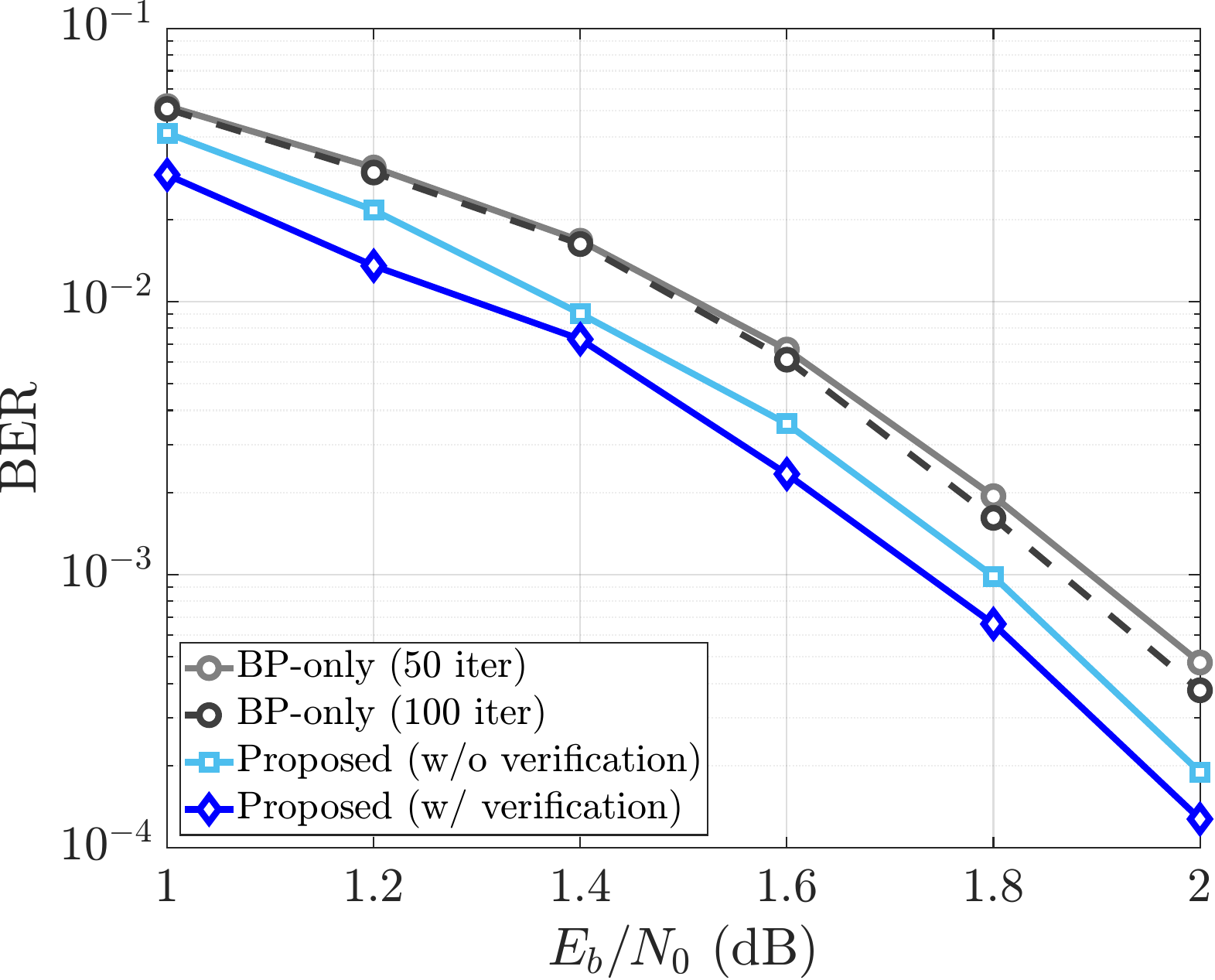}
    \caption{BER versus $E_b/N_0$ for the proposed method and the BP-only baselines.}
    \vspace{-2mm}
    \label{fig:ber}
\end{figure}
\begin{figure} [t]
    \centering
    \includegraphics[width=0.8\linewidth]{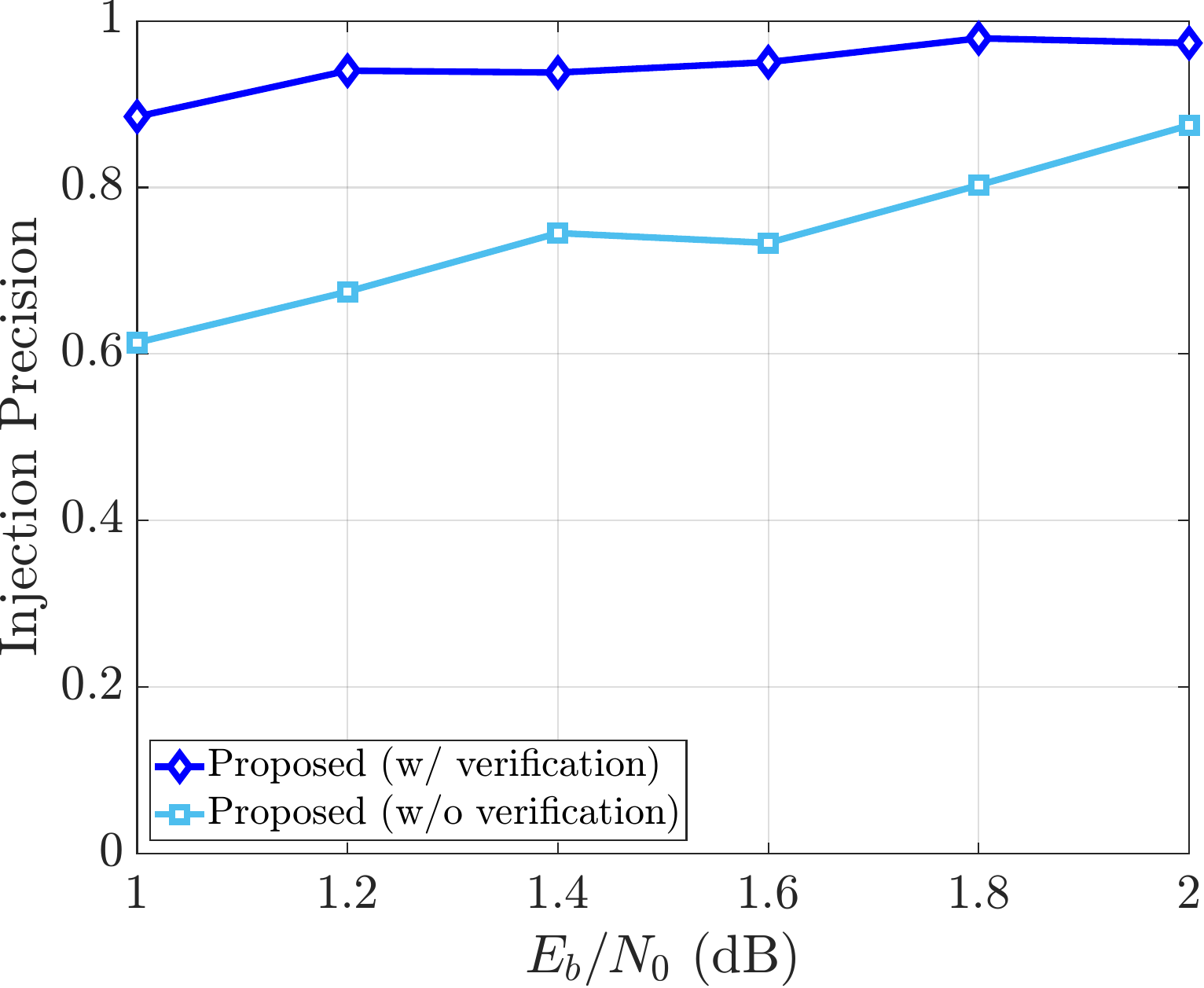}
    \caption{Injection precision versus $E_b/N_0$ for the proposed method with and without verification.}
    \vspace{-2mm}
    \label{fig:precision}
\end{figure}

\section{Simulation Results} \label{sec:sim}
\subsection{Simulation Settings}
\label{sec:setting}

\begin{table}[t]
\centering
\caption{Breakdown of corrected bits into directly injected and propagation-recovered bits under the proposed method with verification.}
\label{tab:propagation}
\begin{tabular}{cccc}
\hline
$E_b/N_0$ (dB) & Direct & Propagated & Prop./Direct \\
\hline
1.0 & 1{,}009 & 4{,}545 & 4.50$\times$ \\
1.2 & 1{,}133 & 4{,}847 & 4.28$\times$ \\
1.4 & 1{,}195 & 4{,}606 & 3.85$\times$ \\
1.6 & 1{,}056 & 5{,}052 & 4.78$\times$ \\
1.8 & 1{,}077 & 4{,}284 & 3.98$\times$ \\
2.0 & 733 & 2{,}848 & 3.89$\times$ \\
\hline
All & 6{,}203 & 26{,}182 & 4.22$\times$ \\
\hline
\end{tabular}
\end{table}

We consider a $(3,6)$-regular systematic LDPC code with $n=1296$ and $k=650$, generated using the pyldpc library~\cite{pyldpc}. The coded bits are modulated with binary phase-shift keying and transmitted over an AWGN channel with noise variance $\sigma^2$, for which the channel LLR in \eqref{eq:llr} reduces to $L_{\mathrm{ch}}(j) = 2y_j/\sigma^2$. Each character is represented by $q=8$ bits, and the text payload is drawn from the Europarl corpus~\cite{koehn-2005-europarl}. GPT-3.5-turbo is employed as the LLM for text correction. The verification tolerance is set to $\epsilon=10$, which corresponds to 1.54\% of the parity checks, and the injection parameters are $\theta=8$ and $\beta=3$.
 
The proposed method operates exclusively on frames exhibiting a nonzero syndrome following the initial decoding pass. To ensure a uniform sample size, simulations at each SNR point continue until exactly $100$ failed frames are accumulated. The decoder initially executes $I_{\max}=50$ iterations. Subsequently, a priori LLRs are injected, and the decoder performs up to $50$ additional iterations, yielding a maximum computational budget of $100$ iterations.

The following schemes are compared:
\begin{itemize}
\item \textbf{BP-only (50 iter):} The conventional decoder with up to $50$ iterations, serving as the primary baseline.
\item \textbf{BP-only (100 iter):} The conventional decoder with up to $100$ iterations, matching the proposed computational budget.
\item \textbf{Proposed (w/o verification):} The proposed method, injecting all LLM corrections without verification.
\item \textbf{Proposed (w/ verification):} The complete proposed method, injecting only the corrections that pass the syndrome-weight verification in \eqref{eq:accept}.
\end{itemize}
 
\subsection{Evaluation Metrics}
\label{sec:metric}
 
The overall decoding performance is evaluated using the bit error rate (BER), defined as the fraction of erroneous bits remaining after the final decoding pass.

To quantify how reliably the injected semantic priors translate into
correct bit decisions, we define the \emph{injection precision}. This metric considers exclusively the bits that receive the prior. We classify such a bit as \emph{fixed} if it transitions from erroneous to correct during the second BP pass. Conversely, it is classified as \emph{broken} if it transitions from correct to erroneous. Letting $N_{\mathrm{fixed}}$ and $N_{\mathrm{broken}}$ denote the total counts for these two cases, the injection precision is defined as
\begin{equation}
P_{\mathrm{inj}} = \frac{N_{\mathrm{fixed}}}
{N_{\mathrm{fixed}} + N_{\mathrm{broken}}}.
\label{eq:precision}
\end{equation}
 
\subsection{Results}
\label{sec:result}

Fig.~\ref{fig:ber} shows the BER of the compared schemes, measured over all transmitted information bits. The proposed method attains the lowest BER across the evaluated SNR range. To separate the impact of the semantic prior injection from the effect of simply running BP twice, we examine the extended baseline. At $2.0$~dB, the BP-only (100 iter) scheme reduces the BER by $21\%$ relative to the BP-only (50 iter) baseline. In contrast, the proposed method with verification yields a $73\%$ reduction. Doubling the iteration budget alone provides only a marginal improvement. Once BP settles into a residual error pattern, additional iterations extract minimal new information from the parity checks. The substantial performance gain originates explicitly from the semantic prior injection.

Notably, the injected characters represent only a minor fraction of the successfully recovered bits. Table~\ref{tab:propagation} highlights this amplification, showing that every directly injected bit recovers an average of 4.2 additional bits during the second BP pass. Instead of the LLM output, message passing over the Tanner graph drives the majority of the corrections.

The verification step plays a crucial role in the proposed architecture. Although the unverified variant outperforms the BP-only baselines by repairing the text, its gain remains consistently lower than the verified method. Injecting raw LLM output introduces false semantic priors, corrupting previously correct bits. Fig.~\ref{fig:precision} illustrates this dynamic. With verification, the injection precision stays between $0.88$ and $0.98$ across the evaluated SNR range, whereas the unverified precision drops to $0.61$ at $1.0$~dB. This gap shows that the verification step maintains a reliable semantic prior despite inaccurate LLM predictions. The benefit is pronounced at low SNR, where the LLM operates on heavily corrupted text. At $1.0$~dB, verification raises the BER reduction from $20\%$ to $44\%$. 
 
\section{Conclusion}
This work introduced an LLM-assisted LDPC decoding framework. The system considers each LLM correction a preliminary suggestion to verify, not a final decision to trust. The framework retains the classical channel code and adds semantic assistance only at the decoder, preserving the separation architecture and its algebraic guarantees. The proposed corrections are tested jointly against the parity-check constraints, and only the accepted ones are injected as soft priors on the channel LLRs. A subsequent BP pass propagates this evidence through the check nodes, recovering on average 4.2 bits for every directly injected bit. Simulations confirm a BER reduction of up to $73\%$ over a conventional decoder at an equal iteration budget, with the injection precision maintained above $0.88$ despite inaccurate LLM predictions. These results demonstrate that semantic information can be reliably incorporated into physical-layer decoding through code-constrained verification.

\bibliographystyle{IEEEtran}
\bibliography{refs}

@IEEEtranBSTCTL{IEEEexample:BSTcontrol,
  CTLuse_forced_etal       = "yes",
  CTLmax_names_forced_etal = "6",
  CTLnames_show_etal       = "1"
}

@article{needleman1970general,
  title={A general method applicable to the search for similarities in the amino acid sequence of two proteins},
  author={Needleman, Saul B and Wunsch, Christian D},
  journal={J. Mol. Biol.},
  volume={48},
  number={3},
  pages={443--453},
  year={1970},
  publisher={Elsevier}
}

@inproceedings{koehn-2005-europarl,
  title={Europarl: A parallel corpus for statistical machine translation},
  author={Koehn, Philipp},
  booktitle={MT summit},
  pages={79--86},
  year={2005}
}

@article{li2026llm,
  title={{LLM-Viterbi}: Semantic-Aware Decoding for Convolutional Codes},
  author={Li, Zhengtong and Yue, Chentao and Hao, Jiafu and Vucetic, Branka and Li, Yonghui},
  journal={arXiv preprint arXiv:2604.19035},
  year={2026}
}

@article{hao2026semantic,
  title={Semantic Error Correction and Decoding for Short Block Codes},
  author={Hao, Jiafu and Yue, Chentao and Liu, Wanchun and Vucetic, Branka and Li, Yonghui},
  journal={arXiv preprint arXiv:2604.22269},
  year={2026}
}

@INPROCEEDINGS{park2026semantic,
  author={Park, Sojeong and Yang, Hyun Jong},
  booktitle={Proc. IEEE Int. Conf. Acoust., Speech, Signal Process. (ICASSP)}, 
  title={Semantic Pilot Design for Data-Aided Channel Estimation Using A Large Language Model}, 
  year={2026},
  volume={},
  number={},
  pages={21531-21535},
  doi={10.1109/ICASSP55912.2026.11462966}}

@misc{pyldpc,
  author       = {Hicham Janati},
  title        = {pyldpc: Simulation of Low Density Parity Check ({LDPC}) Codes},
  howpublished = {\url{https://github.com/hichamjanati/pyldpc}},
  note         = {Version 0.7.9},
  year         = {2020}
}

@ARTICLE{luo2022semantic,
  author={Luo, Xuewen and Chen, Hsiao-Hwa and Guo, Qing},
  journal={IEEE Wireless Commun.}, 
  title={Semantic Communications: Overview, Open Issues, and Future Research Directions}, 
  year={2022},
  volume={29},
  number={1},
  pages={210-219},
  doi={10.1109/MWC.101.2100269}}

@ARTICLE{shao2024theory,
  author={Shao, Yulin and Cao, Qi and Gündüz, Deniz},
  journal={IEEE Trans. Mobile Comput.}, 
  title={A Theory of Semantic Communication}, 
  year={2024},
  volume={23},
  number={12},
  pages={12211-12228},
  doi={10.1109/TMC.2024.3406375}}

@ARTICLE{yang2022semantic,
  author={Yang, Wanting and Du, Hongyang and Liew, Zi Qin and Lim, Wei Yang Bryan and Xiong, Zehui and Niyato, Dusit and Chi, Xuefen and Shen, Xuemin and Miao, Chunyan},
  journal={IEEE Commun. Surveys Tuts.}, 
  title={Semantic Communications for Future Internet: Fundamentals, Applications, and Challenges}, 
  year={2023},
  volume={25},
  number={1},
  pages={213-250},
  doi={10.1109/COMST.2022.3223224}}

@ARTICLE{xie2021deep,
  author={Xie, Huiqiang and Qin, Zhijin and Li, Geoffrey Ye and Juang, Biing-Hwang},
  journal={IEEE Trans. Signal Process.}, 
  title={Deep Learning Enabled Semantic Communication Systems}, 
  year={2021},
  volume={69},
  number={},
  pages={2663-2675},
  doi={10.1109/TSP.2021.3071210}}

@ARTICLE{jiang2024large,
  author={Jiang, Feibo and Dong, Li and Peng, Yubo and Wang, Kezhi and Yang, Kun and Pan, Cunhua and You, Xiaohu},
  journal={IEEE Commun. Mag.}, 
  title={Large {AI} Model Empowered Multimodal Semantic Communications}, 
  year={2025},
  volume={63},
  number={1},
  pages={76-82},
  doi={10.1109/MCOM.001.2300575}}

@ARTICLE{zhao2024lamosc,
  author={Zhao, Yaru and Yue, Yi and Hou, Shoulu and Cheng, Bo and Huang, Yakun},
  journal={IEEE Trans. Cogn. Commun. Netw.}, 
  title={{LaMoSC}: Large Language Model-Driven Semantic Communication System for Visual Transmission}, 
  year={2024},
  volume={10},
  number={6},
  pages={2005-2018},
  doi={10.1109/TCCN.2024.3401712}}

@article{gallager1962low,
  title={Low-density parity-check codes},
  author={Gallager, Robert},
  journal={IRE Trans. Inf. Theory},
  volume={8},
  number={1},
  pages={21--28},
  year={1962},
  publisher={IEEE}
}

@ARTICLE{peng2024robust,
  author={Peng, Xiang and Qin, Zhijin and Tao, Xiaoming and Lu, Jianhua and Hanzo, Lajos},
  journal={IEEE Trans. Wireless Commun.}, 
  title={A Robust Semantic Text Communication System}, 
  year={2024},
  volume={23},
  number={9},
  pages={11372-11385},
  doi={10.1109/TWC.2024.3381950}}

@ARTICLE{huang2022toward,
  author={Huang, Danlan and Gao, Feifei and Tao, Xiaoming and Du, Qiyuan and Lu, Jianhua},
  journal={IEEE J. Sel. Areas Commun.}, 
  title={Toward Semantic Communications: Deep Learning-Based Image Semantic Coding}, 
  year={2023},
  volume={41},
  number={1},
  pages={55-71},
  doi={10.1109/JSAC.2022.3221999}}

@ARTICLE{park2026robust,
  author={Park, Sojeong and Noh, Hyeonho and Yang, Hyun Jong},
  journal={IEEE Trans. Veh. Technol.}, 
  title={Robust Transmission of Punctured Text With Large Language Model-Based Recovery}, 
  year={2026},
  volume={75},
  number={1},
  pages={1737-1742},
  doi={10.1109/TVT.2025.3595593}}

@ARTICLE{thai2026soft,
  author={Thai, Pham Quang},
  journal={IEEE Commun. Lett.}, 
  title={Soft-Information {LLM} Fusion for {LDPC}-Coded Text Over Visible-Light Links}, 
  year={2026},
  volume={30},
  number={},
  pages={1885-1889},
  doi={10.1109/LCOMM.2026.3688465}}

@article{park2026semanticmimo,
  title={Semantic-Aware Data-Aided Channel Estimation with Large Language Models for {MIMO} Systems},
  author={Park, Sojeong and Choi, Jaehyun and Yang, Hyun Jong},
  journal={arXiv preprint arXiv:2607.18640},
  year={2026}
}

@ARTICLE{richardson2018design,
author={Richardson, Tom and Kudekar, Shrinivas},
journal={IEEE Commun. Mag.},
title={Design of Low-Density Parity Check Codes for {5G} New Radio},
year={2018},
volume={56},
number={3},
pages={28-34},
doi={10.1109/MCOM.2018.1700839}}

@techreport{3gpp38212,
author      = {{3GPP}},
title       = {{NR}; Multiplexing and channel coding},
institution = {3rd Generation Partnership Project (3GPP)},
type        = {{TS}},
number      = {38.212},
note        = {Version 15.7.0, Rel. 15},
year        = {2019},
}

@inproceedings{richardson2003error,
author    = {Richardson, Tom},
title     = {Error Floors of {LDPC} Codes},
booktitle = {Proc. 41st Annu. Allerton Conf. Commun., Control, Comput.},
address   = {Monticello, IL, USA},
pages     = {1426--1435},
month     = oct,
year      = {2003}
}

@ARTICLE{dolecek2010analysis,
author={Dolecek, Lara and Zhang, Zhengya and Anantharam, Venkat and Wainwright, Martin J. and Nikolic, Borivoje},
journal={IEEE Trans. Inf. Theory},
title={Analysis of Absorbing Sets and Fully Absorbing Sets of Array-Based {LDPC} Codes},
year={2010},
volume={56},
number={1},
pages={181-201},
doi={10.1109/TIT.2009.2034781}}

@ARTICLE{butler2014error,
author={Butler, Brian K. and Siegel, Paul H.},
journal={IEEE Trans. Inf. Theory},
title={Error Floor Approximation for {LDPC} Codes in the {AWGN} Channel},
year={2014},
volume={60},
number={12},
pages={7416-7441},
doi={10.1109/TIT.2014.2363832}}

@ARTICLE{han2009low,
author={Han, Yang and Ryan, William E.},
journal={IEEE Trans. Commun.},
title={Low-floor decoders for {LDPC} codes},
year={2009},
volume={57},
number={6},
pages={1663-1673},
doi={10.1109/TCOMM.2009.06.070325}}

@ARTICLE{kang2011iterative,
author={Kang, Jingyu and Huang, Qin and Lin, Shu and Abdel-Ghaffar, Khaled},
journal={IEEE Trans. Commun.},
title={An Iterative Decoding Algorithm with Backtracking to Lower the Error-Floors of {LDPC} Codes},
year={2011},
volume={59},
number={1},
pages={64-73},
doi={10.1109/TCOMM.2010.101210.090628}}

\end{document}